\documentclass[twocolumn,showpacs,amsmath,amssymb,aps,prl]{revtex4}
\usepackage{dcolumn}
\usepackage{bm}

\def\res{C_{\rm res}}
\newcommand{\n}{\mbox{\boldmath $\nabla$}}

\usepackage[dvips]{graphicx}
\usepackage{wrapfig,subfigure}
\usepackage{amssymb}
\usepackage{color}

\begin{document}

\title{Critical exponents in metastable decay via quantum activation}
\author{M. I. Dykman}
\affiliation{Department of Physics
and Astronomy, Michigan State
University, East Lansing, MI 48824, USA}
\date{\today}

\begin{abstract}
We consider decay of metastable states of forced vibrations of a
quantum oscillator close to bifurcation points, where dissipation
becomes effectively strong. We show that decay occurs via quantum
activation over an effective barrier. The decay probability $W$
scales with the distance $\eta$ to the bifurcation point as $|\ln
W|\propto \eta^{\xi}$. The exponent $\xi$ is found for a resonantly
driven oscillator and an oscillator modulated at nearly twice its
eigenfrequency.
%
\end{abstract}

\maketitle

Decay of a metastable state is usually considered as resulting from
tunneling or thermal activation. In this paper we study a different
decay mechanism, quantum activation. It relates to systems far from
thermal equilibrium. As tunneling, quantum activation is due to
quantum fluctuations, but as thermal activation, it involves
diffusion over an effective barrier separating the metastable state.

Metastable decay in nonequilibrium systems has attracted much
attention recently in the context of switching between coexisting
states of forced vibrations. Such diverse systems as trapped
electrons and atoms \cite{Lapidus1999,Kim2005}, Josephson junctions
\cite{Siddiqi2005,Siddiqi2005a}, and nano- and micromechanical
oscillators \cite{Aldridge2005,Stambaugh2005} have been studied. The
experiments largely focused on the parameter range where the system
was close to a bifurcation point in which the metastable state
disappears. In this range the decay probability is comparatively
large and displays characteristic scaling with the distance to the
bifurcation point. So far classical activation was studied, but
recently quantum regime has been also reached
\cite{Devoret_private06}.

For classical systems, scaling of the rate of activated decay near a
bifurcation point was found theoretically both in the cases of
equilibrium \cite{Kurkijarvi1972,Victora1989,Garg1995a} and
nonequilibrium systems
\cite{Dykman1980,Dmitriev1986a,Tretiakov2005}. In the latter case a
scaling crossover may occur as the system goes from the underdamped
to overdamped regime while approaching the bifurcation point
\cite{Dykman2005b}. Such crossover occurs also for quantum tunneling
in equilibrium dissipative systems \cite{Caldeira1983}.

In this paper we study decay of metastable vibrational states in
dissipative systems close to bifurcation points, where the motion
becomes overdamped. The analysis refers to the systems of current
interest, quantum oscillators driven by a resonant force or
parametrically modulated at nearly twice the eigenfrequency. We show
that at low temperatures decay occurs via quantum activation. The
decay rate $W$ scales with the distance to the bifurcation point
$\eta$ as $|\ln W|\propto\eta^{\xi}$. The scaling exponent is
$\xi=3/2$ for resonant driving, and $\xi=2$ for parametric
modulation; in addition, $|\ln W|$ displays a characteristic
temperature dependence.

Quantum activation in periodically modulated systems can be
understood by noting that metastable states are formed as a result
of the balance between external driving and dissipation due to
coupling to a thermal bath. For $T=0$ dissipation corresponds to
transitions to lower energy states with emission of excitations of
the bath. However, modulated systems are more adequately described
by the Floquet (quasienergy) states than by the energy eigenstates.
Emission of bath excitations may result in transitions to both
higher and lower quasienergies, albeit with different probabilities
\cite{Dykman1988a,Marthaler2006}. The higher-probability transitions
lead to relaxation towards a metastable state, whereas the
lower-probability transitions lead to effective diffusion away from
it, a finite-width distribution over quasienergy, and metastable
decay. There is certain similarity here with the Unruh effect
\cite{Unruh1976} where a uniformly accelerated relativistic detector
coupled to a quantum zero-temperature field is described in its
proper time by the Gibbs distribution with the
acceleration-dependent temperature.

We will start with a resonantly driven nonlinear oscillator. Its
Hamiltonian is
\begin{equation}
\label{eq:Hamiltonian_res} H_0(t)=\frac{1}{2}p^2 +
\frac{1}{2}\omega_0^2q^2 + \frac{1}{4}\gamma q^4 -qA\cos(\omega_Ft).
\end{equation}
In the presence of weak damping the oscillator may have two
coexisting stable states of classical forced vibrations
\cite{LL_Mechanics2004}. They emerge already for a small modulation
amplitude $A$ provided the detuning $\delta\omega = \omega_F -
\omega_0$ of the modulation frequency $\omega_F$ from the oscillator
eigenfrequency $\omega_0$ is small, $|\delta\omega|\ll \omega_F$. We
assume that the nonlinearity is small, $|\gamma|\langle q^2\rangle
\ll \omega_0^2$, and that $\gamma\,\delta\omega > 0$, which is
necessary for the onset of bistability.

It is convenient to switch from $q,p$ to slowly varying operators
$Q,P$, using a transformation $q=\res(Q\cos\omega_F
t+P\sin\omega_F t)$, $p=-\res\omega_F (Q\sin\omega_F t -
P\cos\omega_F t)$ with $\res=(8\omega_F\delta\omega/3\gamma)^{1/2}$.
The variables $Q,P$ are the scaled coordinate and momentum in the
rotating frame,
\begin{equation}
\label{eq:commutator} [P,Q]=-i\lambda,\qquad \lambda
=3\hbar\gamma/8\omega_F^2\,\delta\omega.
\end{equation}
The parameter $\lambda$ plays the role of the effective Planck
constant. We are interested in the semiclassical case; $\lambda$ is
the small parameter of the theory, $\lambda \ll 1$.

In the rotating wave approximation the Hamiltonian
(\ref{eq:Hamiltonian_res}) for $\delta\omega > 0$ becomes $ H_0=
(\hbar/\lambda)\delta\omega\,\hat g$, with
\begin{eqnarray}
\label{eq:g_resonant}
\hat g\equiv g(Q,P)&=&\frac{1}{4}(Q^2+P^2-1)^2 - \beta^{1/2}Q, \\
\beta&=&3\gamma
A^2/32\omega_F^3\left(\delta\omega\right)^3.\nonumber
\end{eqnarray}
(for $\delta\omega <0$ one should redefine $g\to -g, H_0\to
-(\hbar/\lambda)\delta\omega\,g $). The function $g$ plays the role
of the oscillator Hamiltonian in dimensionless time $\tau =
t|\delta\omega|$. The eigenvalues of $g$ give oscillator
quasienergies.

The parameter $\beta$ in Eq.~(\ref{eq:g_resonant}) is the scaled
intensity of the driving field. For weak damping the oscillator is
bistable provided $0<\beta < 4/27$. In this range the function
$g(Q,P)$ has a shape of a tilted Mexican hat. The maximum at the top
of the central dome and the minimum at the lowest point of the rim
correspond, respectively, to the small- and large-amplitude states
of forced vibrations. The saddle point of $g$ corresponds to the
unstable periodic state of the oscillator.

We will consider two major relaxation mechanisms of the oscillator:
damping due to coupling to a thermal bath and dephasing due to
oscillator frequency modulation by an external noise. Usually the
most important damping mechanism is transitions between neighboring
oscillator energy levels. They result from the coupling linear in
the oscillator coordinate. Since the energy transfer is $\approx
\hbar\omega_0$, in the rotating frame the transitions look
instantaneous. We will assume that the correlation time of the noise
that modulates the oscillator frequency is also short compared to
$1/|\delta\omega|$, so that the noise is effectively
$\delta$-correlated in slow time $\tau$. Then the quantum kinetic
equation is Markovian in the rotating frame,
\begin{eqnarray}
\label{eq:QKE_general} \dot\rho\equiv
\partial_{\tau}\rho=i\lambda^{-1}[\rho,g]-\hat{\Gamma}\rho\ -
\hat{\Gamma}^{\rm ph}\rho,
\end{eqnarray}
where $\hat\Gamma\rho$ describes damping
\begin{eqnarray}
\label{eq:decay}
\hat{\Gamma}\rho&=&\Gamma|\delta\omega|^{-1}\left[(\bar{n}+1)(\hat
a^{\dagger}\hat a\rho-2\hat
a\rho\hat a^{\dagger}+ \rho\hat a^{\dagger}\hat a)\right.\nonumber\\
&&\left.+\bar{n}(\hat a\hat a^{\dagger}\rho-2\hat
a^{\dagger}\rho\hat a +\rho\hat a\hat a^{\dagger})\right],
\end{eqnarray}
and $\hat{\Gamma}^{\rm ph}\rho$ describes dephasing,
\begin{equation}
\label{eq:dephasing} \hat{\Gamma}^{\rm ph}\rho=\Gamma^{\rm
ph}|\delta\omega|^{-1}\left[\hat a^{\dagger}\hat a,\left[\hat
a^{\dagger}\hat a,\rho\right]\right].
\end{equation}
Here, $\Gamma$ and $\Gamma^{\rm ph}$ are the damping and dephasing
rates, $\hat a= (2\lambda)^{-1/2}(Q+iP)$ is the lowering operator,
and $\bar n = [\exp\left(\hbar\omega_0/k T\right)-1]^{-1}$ is the
oscillator Planck number. In what follows we use dimensionless
parameters
\begin{equation}
\label{eq:rates_ratio}\Omega=|\delta\omega|/\Gamma,\qquad
\varkappa^{\rm ph} =\Gamma^{\rm ph}/\lambda\Gamma.
\end{equation}
We assume that $\varkappa^{\rm ph}\lesssim 1$. This means that the
dephasing fluctuations intensity may be comparable to the intensity
of quantum fluctuations associated with damping, which is $\propto
\lambda\Gamma$, see below, but that $\Gamma^{\rm ph}\ll \Gamma$.

Metastable decay of the driven oscillator was studied earlier
\cite{Dykman1988a} assuming that the damping-induced broadening of
quasienergy levels is small compared to the typical interlevel
distance. This condition necessarily breaks near a bifurcation point
where local extrema of $g(Q,P)$ come close to each other and the
motion is slowed down. Therefore the analysis should be done
differently. It is simplified in the Wigner representation of the
density matrix,
\begin{eqnarray}
\label{eq:Wigner_def} \rho_W( Q,P)=\int d\xi e^{-i\xi
P/\lambda}\rho\left( Q+\frac{1}{2}\xi, Q-\frac{1}{2}\xi\right),
\end{eqnarray}
where $\rho(Q_1,Q_2)=\langle Q_1|\rho| Q_2\rangle$ is the density
matrix in the coordinate representation. Using
Eqs.~(\ref{eq:commutator})-(\ref{eq:Wigner_def}) one can formally
write the equation for $\rho_W$ as a sum of terms proportional to
different powers of $\lambda$,
\begin{eqnarray}
\label{eq:Liouville_compact} \dot\rho_W=-\n\left({\bf
K}\rho_W\right) +\lambda \hat L^{(1)}\rho_W +\lambda^2\hat
L^{(2)}\rho_W.
\end{eqnarray}
Here we introduced vectors ${\bf K} = (K_Q,K_P)$ and $\n =
(\partial_Q,\partial_P)$.

Vector ${\bf K}$ in Eq.~(\ref{eq:Liouville_compact}) determines the
evolution of the density matrix in the absence of quantum and
classical fluctuations,
\begin{eqnarray}
\label{eq:K_vector} K_Q=\partial_P g-\Omega^{-1}Q \qquad
K_P=-\partial_P g-\Omega^{-1}P.
\end{eqnarray}
This evolution corresponds to classical motion
\begin{equation}
\label{eq:classical} \dot Q=K_Q,\qquad\dot P=K_P.
\end{equation}
The condition ${\bf K}={\bf 0}$ gives the values of $Q, P$ at the
stationary states of the oscillator in the rotating frame.

The term $\hat L^{(1)}$ in Eq.~(\ref{eq:Liouville_compact})
describes classical and quantum fluctuations due to damping and
dephasing,
\begin{equation}
\label{eq:L1} \hat L^{(1)} = \Omega^{-1}\left[\left(\bar
n+\frac{1}{2}\right)\n^2 + \varkappa^{\rm
ph}\left(Q\partial_P-P\partial_Q\right)^2\right].
\end{equation}
These fluctuations lead to diffusion in $(Q,P)$-space, as seen from
the structure of $\hat L^{(1)}$.

The term $\hat L^{(2)}$ in Eq.~(\ref{eq:Liouville_compact})
describes quantum effects of motion of the isolated oscillator,
\begin{equation}
\label{eq:L2} \hat L^{(2)} =
-\frac{1}{4}\left(Q\partial_P-P\partial_Q\right)\n^2.
\end{equation}
In contrast to $\hat L^{(1)}$, the operator $\hat L^{(2)}$ contains
third derivatives. Generally the term $\lambda^2\hat L^{(2)}\rho_W$
is not small, because $\rho_W$ varies on distances $\sim\lambda$.
However, it becomes small close to bifurcation points, as shown
below.

From Eqs.~(\ref{eq:K_vector}), (\ref{eq:classical}), for given
damping $\Omega^{-1}$ the oscillator has two stable and one unstable
stationary state in the rotating frame (periodic states of forced
vibrations) in the range $\beta_B^{(1)}(\Omega) < \beta <
\beta_B^{(2)}(\Omega)$ and one stable state outside this range
\cite{LL_Mechanics2004}, with
\begin{equation} \label{eq:beta_B}
\beta_B^{(1,2)}=\frac{2}{27}\left[1+9\Omega^{-2}
\mp\left(1-3\Omega^{-2}\right)^{3/2}\right].
\end{equation}
At $\beta_B^{(1)}$ and $\beta_B^{(2)}$ the stable states with large
and small $Q^2+P^2$, respectively (large and small vibration
amplitudes), merge with the saddle state (saddle-node bifurcation).
The values of $Q,P$ at the bifurcation points 1, 2 are
$Q_B=\beta_B^{-1/2}Y_B(Y_B-1)$, $P_B=\beta_B^{-1/2}\Omega^{-1}Y_B$,
where $Y_B=Q_B^2+P_B^2$,
\begin{eqnarray}
\label{eq:positions_bif}
Y_B^{(1,2)}=\frac{1}{3}\left[2\pm(1-3\Omega^{-2})^{1/2}\right].
\end{eqnarray}

In the absence of fluctuations dynamics of a classical system near a
saddle-node bifurcation point is controlled by one slow variable
\cite{Guckenheimer1987}. In our case it can be found by expanding
$K_{Q,P}$ in $\delta Q= Q-Q_B, \,\delta P=P-P_B$, and the distance
to the bifurcation point $\eta=\beta-\beta_B$. The function $K_P$
does not contain linear terms in $\delta Q, \delta P$. Then, from
Eq.~(\ref{eq:classical}), $P$ slowly varies in time for small
$\delta Q, \delta P, \eta$. On the other hand
\begin{eqnarray}
\label{eq:K_Q_B} K_Q\approx -2\Omega^{-1}\left(\delta Q- a_B\delta
P\right),\quad a_B=\Omega(2Y_B-1).
\end{eqnarray}
Therefore the relaxation time of $Q$ is $\Omega/2$, it does not
depend on the distance to the bifurcation point. As a consequence,
$Q$ follows $P$ adiabatically, i.e., over time $\sim\Omega$ it
adjusts to the instantaneous value of $P$.

The adiabatic approximation can be applied also to fluctuating
systems. The approach is well known for classical systems described
by the Fokker-Planck equation \cite{Haken2004}. We now extend it to
the quantum problem.

Formally we change in Eq.~(\ref{eq:Liouville_compact}) from $Q$ and
$P$ to $\delta\tilde Q= \delta Q-a_B\delta P$ and $\delta P$. For
times $\tau\gg \Omega^{-1}$ the distribution $\rho _W$ has a narrow
peak as a function of $\delta\tilde Q$, whereas its dependence on
$\delta P$ is much more smooth. We seek $\rho_W$ near its maximum
over $\delta\tilde Q$ in the form
\begin{eqnarray}
\label{eq:adiabatic_form}
 \rho_W=(2\pi \lambda\sigma^2)^{-1/2}
 \exp\left(-\delta\tilde Q^2/2\lambda\sigma^2\right)\bar\rho_W(\delta P),
\end{eqnarray}
where
$\sigma^2=\frac{1}{2}(1+a_B^2)\left(\bar
n+\frac{1}{2}\right)+\frac{1}{8}\varkappa^{\rm ph}\beta_B\Omega^2.
$
The $\delta\tilde{Q}$-dependent factor in $\rho_W$ is chosen so that
in Eq.~(\ref{eq:Liouville_compact}) the term
$\partial_{\delta\tilde{Q}}K_Q\rho_W$ and the term $\propto
\lambda\partial_{\delta\tilde{Q}}^2\rho_W$ compensate each other.
Note that corrections from $\lambda^2\hat L^{(2)}\rho_W$ are of
higher order in $\lambda$ for $\delta\tilde{Q}^2 \lesssim \lambda$.

The function $\bar\rho_W$ describes the distribution over $\delta
P$. In the spirit of the adiabatic approximation, it can be
calculated disregarding small fluctuations of $Q$, i.e., setting
$\delta\tilde{Q}=0$ in Eq.~(\ref{eq:Liouville_compact}). Formally,
one obtains an equation for $\bar\rho_W$ by substituting
Eq.~(\ref{eq:adiabatic_form}) into the full kinetic equation
(\ref{eq:Liouville_compact}) and integrating over $\delta\tilde{Q}$.
This gives
\begin{eqnarray}
\label{eq:eq_bar_rho} \dot{\bar\rho}_W\approx
\partial_P\left[\bar\rho_W\partial_PU
+\lambda {\cal D}_B\partial_P\bar\rho_W \right],
\end{eqnarray}
where $U$ and ${\cal D}$ have the form
\begin{eqnarray}
\label{eq:U(x)_D} &&U=\frac{1}{3}b(\delta P)^3
-\frac{1}{2}\beta_B^{-1/2}\eta\delta P, \qquad \eta=\beta-\beta_B,\nonumber\\
&&{\cal D}_B=\Omega^{-1}\left[\left(\bar n+\frac{1}{2}\right)
+\frac{1}{2}\varkappa^{\rm ph}(1-Y_B)\right]
\end{eqnarray}
with $b=-\beta_B^{1/2}(2Y_B)^{-1}(1-2\Omega^2Y_B+\Omega^2)$. In
Eqs.~(\ref{eq:eq_bar_rho}), (\ref{eq:U(x)_D}) we kept only the
lowest order terms in $\delta P, \beta-\beta_B,\lambda$. In
particular we dropped the term
$-\lambda^2Q_B\partial_P^3\bar\rho_W/4$ which comes from the
operator $\hat L^{(2)}$ in Eq.~(\ref{eq:Liouville_compact}). One can
show that, for typical $|\delta P|\sim |\eta|^{1/2}$, this term
leads to corrections $\sim \eta,\lambda$ to $\bar\rho_W$.

Eq.~(\ref{eq:eq_bar_rho}) has a standard form of the equation for
classical diffusion in a potential $U(\delta P)$, with diffusion
coefficient $\lambda{\cal D}_B$. For $\eta b>0$ the potential $U$
has a minimum and a maximum. They correspond to the stable and
saddle states of the oscillator. The distribution $\rho_W$ has a
diffusion-broadened peak at the stable state. Diffusion also leads
to escape from the stable state, i.e., to metastable decay. The
decay rate $W$ is given by the Kramers theory \cite{Kramers1940},
\begin{eqnarray}
\label{eq:W_res} W=Ce^{-R_A/\lambda},\qquad
R_A=\frac{2^{1/2}|\eta|^{3/2}}{3{\cal D}_B|b|^{1/2}\beta_B^{3/4}},
\end{eqnarray}
with prefactor
$C=\pi^{-1}(b\eta/2)^{1/2}\beta_B^{-1/4}|\delta\omega|$ (in unscaled
time $t$).

The rate (\ref{eq:W_res}) displays activation dependence on the
effective Planck constant $\lambda$.  The characteristic quantum
activation energy $R_A$ scales with the distance to the bifurcation
point $\eta=\beta-\beta_B$ as $\eta^{3/2}$. This scaling is
independent of temperature. However, the factor ${\cal D}_B$ in
$R_A$ displays a characteristic $T$ dependence. In the absence of
dephasing we have ${\cal D}_B=1/2\Omega$ for $\bar n\ll 1$, whereas
${\cal D}_B= kT/\hbar\omega_0\Omega$ for $\bar n\gg 1$. In the
latter case the expression for $W$ coincides with the result
\cite{Dykman1980}.

In the limit $\Omega\gg 1$ the activation energy (\ref{eq:W_res})
for the small-amplitude state has the same form as in the range of
$\beta$ still close but further away from the bifurcation point,
where the distance between quasienergy levels largely exceeds their
width \cite{Dykman1988a}. We note that the rate of tunneling decay
for this state is exponentially smaller; the tunneling exponent for
constant quasienergy scales as $\eta^{5/4}$ \cite{Dmitriev1986a},
which is parametrically larger than $\eta^{3/2}$ for small $\eta$
[for comparison, for a particle in a cubic potential
(\ref{eq:U(x)_D}) the tunneling exponent in the strong-damping limit
scales as $\eta$ \cite{Caldeira1983}].

For the large-amplitude state the quantum activation energy,
Eq.~(\ref{eq:W_res}), displays different scaling from that further
away from the bifurcation point, where $R_A\propto\beta^{1/2}$ for
$\Omega \gg 1$ \cite{Dykman1988a}. For this state we therefore
expect a scaling crossover to occur with varying $\beta$.

The approach to decay of vibrational states can be extended to a
parametrically modulated oscillator. The Hamiltonian of such an
oscillator is
\begin{equation}
\label{eq:H_0_param(t)}
H_0(t)=\frac{1}{2}p^2+\frac{1}{2}q^2\left[\omega_0^2+F\cos(\omega_F
t)\right]+\frac{1}{4}\gamma q^4\, .
\end{equation}
When the modulation frequency $\omega_F$ is close to $2\omega_0$, as
a result of parametric resonance the oscillator may have two stable
states of vibrations at frequency $\omega_F/2$ (period-two states)
shifted in phase by $\pi$ \cite{LL_Mechanics2004}. For $F\ll
\omega_0^2$ the oscillator dynamics is characterized by the
dimensionless frequency detuning $\mu$, effective Planck constant
$\lambda$, and relaxation time $\zeta$,
\begin{equation}
\label{eq:mu_and_lambda} \mu =
\frac{\omega_F(\omega_F-2\omega_0)}{F}, \quad \lambda =
\frac{3|\gamma|\hbar}{F\omega_F}, \quad
\zeta=\frac{F}{2\omega_F\Gamma}.
\end{equation}
As before, $\lambda$ will be the small parameter of the theory.

Parametric excitation requires that the modulation be sufficiently
strong, $\zeta > 1$. For such $\zeta$ the bifurcation values of
$\mu$ are
\begin{equation}
\label{eq:mu_B} \mu_B^{(1,2)}= \mp (1-\zeta^{-2})^{1/2}, \qquad
\zeta
> 1.
\end{equation}
If $\gamma>0$, as we assume, for $\mu<\mu_B^{(1)}$ the oscillator
has one stable state; the vibration amplitude is zero. As $\mu$
increases and reaches $\mu_B^{(1)}$ this state becomes unstable and
there emerge two stable period two states (a supercritical pitchfork
bifurcation). They remain stable for larger $\mu$. In addition, when
$\mu$ reaches $\mu_B^{(2)}$ the zero-amplitude state also becomes
stable (a subcritical pitchfork bifurcation). The case $\gamma < 0$
is described by replacing $\mu\to -\mu$.

The classical fluctuation-free dynamics for $\mu$ close to $\mu_B$
is controlled by one slow variable \cite{Guckenheimer1987}. The
analysis analogous to that for the resonant case shows that, in the
Wigner representation, fluctuations are described by one-dimensional
diffusion in a potential, which in the present case is quartic in
the slow variable. The probability $W$ of switching between the
period-two states for small $\mu - \mu_B^{(1)}$ and the decay
probability of the zero-amplitude state for small $\mu -
\mu_B^{(2)}$ have the form $W=C\exp(-R_A/\lambda)$ with
\begin{eqnarray}
\label{eq:W_par} R_A=|\mu_B|\eta^2/2(2\bar n+1), \qquad \eta=\mu -
\mu_B
\end{eqnarray}
($\mu_B=\mu_B^{(1,2)}$). The corresponding prefactors are
$C_B^{(2)}=2C_B^{(1)}=2^{1/2}\pi^{-1}\Gamma\zeta^2|\mu_B||\mu -
\mu_B|$. We note that dephasing does not affect the decay rate, to
zeroth order in $\mu-\mu_B$.

From Eq.~(\ref{eq:W_par}), at parametric resonance the quantum
activation energy $R_A$ scales with the distance to the bifurcation
point as $\eta^2$. In the limit $\zeta \gg 1$ the same expression as
Eq.~(\ref{eq:W_par}) describes switching between period-two states
still close but further away from the bifurcation point, where the
distance between quasienergy levels largely exceeds their width. The
exponent for tunneling decay in this case scales as $\eta$
\cite{Marthaler2006}.

It follows from the above results that, both for resonant and
parametric modulation, close to bifurcation points decay of
metastable vibrational states occurs via quantum activation. It
results from diffusion over a barrier. The quantum activation energy
is smaller than the tunneling exponent. Near bifurcation points
these quantities become parametrically different and scale as
different powers of the distance to the bifurcation point.

The exponent of the decay rate displays a characteristic dependence
on temperature. In the absence of dephasing, for $kT\gg
\hbar\omega_0$ we have standard thermal activation, $R_A\propto
1/T$. The low-temperature limit is described by the same expression
with $kT$ replaced by $\hbar\omega_0/2$. Quantum activation imposes
a limit on the sensitivity of bifurcation amplifiers based on
modulated Josephson oscillators used for quantum measurements
\cite{Siddiqi2005,Siddiqi2005a}.

In conclusion, we have studied decay of metastable states of forced
vibrations of a quantum oscillator. Both energy dissipation from
coupling to a bath and noise-induced dephasing were taken into
account. We have found the exponent and the prefactor in the decay
rate near bifurcation points. The quantum activation energy for
resonantly excited period one states scales with the distance $\eta$
to the bifurcation point as $\eta^{3/2}$, whereas for parametrically
excited period two states it scales as $\eta^2$.

I am grateful to M. Devoret for the discussion and for pointing out
the analogy between quantum activation and the Unruh effect. This
research was supported in part  by the NSF through grant No.
PHY-0555346.


\end{document}